\RequirePackage{fix-cm}
\documentclass[twocolumn]{svjour3}          
\smartqed  

\usepackage{graphicx}
\usepackage[numbers,sort&compress]{natbib}
\usepackage{xcolor}
\usepackage{hyperref}
\hypersetup{
    colorlinks,
    linkcolor={black!50!black},
    citecolor={blue!50!black},
    urlcolor={blue!80!black}
}

\usepackage{lineno}
\usepackage{soul}
\usepackage{lscape}
\usepackage{booktabs}
\usepackage{amsmath,mathtools}
\usepackage[intoc]{nomencl}
\usepackage{url}
\usepackage{doi}
\usepackage[boxruled,vlined,linesnumbered]{algorithm2e}
\usepackage{amsfonts}
\usepackage{relsize}
\usepackage{commath}
\usepackage{float}
\usepackage{textcomp}
\usepackage{array}
\usepackage[nottoc,notlot,notlof]{tocbibind}
\usepackage{setspace}
\usepackage{mathptmx}

\usepackage{subcaption}
\graphicspath{{plots/}}

\begin{document}

\title{Use of low-fidelity models with machine-learning error correction for well placement optimization}
\author{Haoyu Tang \and Louis J.~Durlofsky}

\institute{Haoyu Tang \at
              Department of Energy Resources Engineering, \\
              Stanford University, Stanford CA 94305, USA \\
              \email{hytang@stanford.edu}           
           \and
           Louis J. Durlofsky \at
              Department of Energy Resources Engineering, \\
              Stanford University, Stanford CA 94305, USA \\
              \email{lou@stanford.edu}
}

\date{Received: date / Accepted: date}
\maketitle

\begin{abstract}
Well placement optimization is commonly performed using population-based global stochastic search algorithms. These optimizations are computationally expensive due to the large number of multiphase flow simulations that must be conducted. In this work, we present an optimization framework in which these simulations are performed with low-fidelity (LF) models. These LF models are constructed from the underlying high-fidelity (HF) geomodel using a global transmissibility upscaling procedure. Tree-based machine-learning methods, specifically random forest and light gradient boosting machine, are applied to estimate the error in objective function value (in this case net present value, NPV) associated with the LF models. In the offline (preprocessing) step,  preliminary optimizations are performed using LF models, and a clustering procedure is applied to select a representative set of 100--150 well configurations to use for training. HF simulation is then performed for these configurations, and the tree-based models are trained using an appropriate set of features. In the online (runtime) step, optimization with LF models, with the machine-learning correction, is conducted. Differential evolution is used for all optimizations. Results are presented for two example cases involving the placement of vertical wells in 3D bimodal channelized geomodels. We compare the performance of our procedure to optimization using HF models. In the first case, 25 optimization runs are performed with both approaches. Our method provides an overall speedup factor of 46 relative to optimization using HF models, with the best-case NPV within 1\% of the HF result. In the second case fewer HF optimization runs are conducted (consistent with actual practice), and the overall speedup factor with our approach is about 8. In this case, the best-case NPV from our procedure exceeds the HF result by 3.8\%.

\end{abstract}

\section{Introduction}

The efficient management of subsurface flow operations such as oil/gas production, geological CO$_2$ storage and aquifer remediation requires the solution of challenging nonlinear optimization problems. The general problem is very involved and can entail optimizing the drilling schedule, well locations and time-varying settings for each well. Often, the most fundamental issue is the determination of the optimal well locations. This well placement optimization (WPO) problem, which is the subject of this study, is characterized by many local optima and is typically addressed in practice using population-based global stochastic search algorithms. These techniques are expensive because many function evaluations are required, and each of these may involve a detailed multiphase flow simulation. Thus, procedures that act to accelerate the optimizations can have a great impact in practical settings.

In this paper, we develop an optimization treatment that entails the use of low-fidelity flow simulation models applied in conjunction with machine-learning-based error models. Our target application is an oil production problem in which we seek to maximize the net present value (NPV) of the project. The low-fidelity (LF) models are constructed from the underlying high-fidelity (HF) 3D geological model using a formal flow-based upscaling, or grid coarsening, procedure. The machine-learning (ML) error model uses two tree-based techniques -- random forest and light gradient boosting machine -- for estimating the error in LF objective function values relative to those from HF flow simulation. This requires performing 100--150 HF simulation runs in an offline preprocessing step, which is many fewer simulations than are typically needed in deep-learning-based procedures. Optimization in this work is achieved using a differential evolution algorithm. The overall methodology is applied to determine the optimal locations for vertical injection and production wells in heterogeneous 3D geomodels.

Many investigators, using a range of optimization algorithms, have addressed general problems in field development optimization (e.g., \citep{li2012variable}, \citep{forouzanfar2014joint}, \citep{isebor2014compgeo}, \citep{forouzanfar2016simultaneous}, \citep{jahandideh2020closed}, \citep{de2021field}) and in stand-alone well placement optimization (e.g., \citep{nasrabadi2012well}, \citep{onwunalu2010application}, \citep{volkov2018gradient}). Because these optimizations can be very time consuming, considerable effort has been directed toward the development of surrogate (or proxy) models to accelerate the flow simulations. Surrogate methods can often be classified as either reduced-physics/reduced-numerics methods, or as data-driven methods. The former category includes the widely studied reduced-order models based on proper orthogonal decomposition (POD). POD-based approaches were first applied for well control optimization, in which optimal time-varying well settings are determined, by van Doren et al.~\citep{van2006reduced}. More recent work in this area is discussed in the review by Jansen and Durlofsky \citep{jansen2017use}. POD-based methods have received limited attention for WPO problems, though they were applied in this setting by Zalavadia and Gildin \citep{zalavadia2018parametric, zalavadia2021two}. Other simplified modeling procedures, such as capacitance-resistance methods, have also been used for well control optimization \citep{sayarpour2009use}.

Another means for accelerating optimizations in reservoir engineering settings is to perform the majority of the flow simulations using upscaled (LF) models. Because the underlying HF geological models often display heterogeneous features that vary on multiple length scales, the upscaling problem must be addressed carefully. Aliyev and Durlofsky \citep{aliyev2015multilevel,aliyev2017multilevel} developed a multifidelity WPO strategy, in which optimization was performed sequentially on models of varying levels of resolution. These models were constructed using global flow-based transmissibility upscaling procedures \citep{zhang2008new, chen2008nonlinear}.
Krogstad et al.~\citep{krogstad2016reservoir} also successfully used upscaled models for well control optimization. These treatments share some similarities with approaches in other application areas. Kontogiannis and Savill \citep{kontogiannis2020generalized}, for example, used LF and HF models in aerostructural design problems, and Korondi et al.~\citep{korondi2021multi} developed a surrogate-assisted strategy using multifidelity models for design optimization under uncertainty.

Data-driven approaches entail an offline stage, in which some number of HF simulations are performed and the surrogate model is trained, and an online (runtime) stage, where the surrogate model is used to perform function evaluations in the target optimization problem. Data-driven methods have been applied for well control optimization and for WPO, and a range of machine-learning methods have been considered. For example, Golzari et al.~\citep{golzari2015development} used shallow artificial neural networks, while Kim and Durlofsky \citep{kim2021recurrent} applied deep recurrent neural networks, for well control optimization. In WPO settings, Nwachukwu et al.~\citep{nwachukwu2018fast} applied the gradient boosting (GB) method for a range of optimization problems involving 2D reservoir models. They reported that their machine-learning model benefited from the use of features related to geological connectivity. Nwachukwu et al.~\citep{nwachukwu2018fast} required 500--1000 training runs to accurately predict NPV in 2D models containing $50 \times 50$ cells.

Kim et al.~\citep{kim2020robust} developed a deep-learning method involving convolutional neural networks (CNNs) for WPO problems. They found that time-of-flight maps, which quantify fluid travel time along streamlines, represented useful inputs to their CNN model. These quantities are not typically computed by reservoir simulators. Kim et al.~\citep{kim2020robust} applied their method for 2D and 3D waterflooding cases, with geological uncertainty included. Training of their deep-learning method required about 4000 HF simulations, and they reported an overall speedup factor of 5.2. Wang et al.~\citep{wang2021efficient} applied a theory-guided CNN (TgCNN) procedure for WPO with uncertainty. TgCNN incorporates physical constraints, which derive from the partial differential equations that describe the flow problem, during the training process. They performed optimization in 2D systems under single-phase flow.
Although the TgCNN runtime cost is very low, training is expensive, and overall speedup with the current implementation was modest.


The surrogate treatment developed in this work includes both reduced-numerics and data-driven components. During runtime optimizations, we perform flow simulation on LF (highly upscaled) models. These models are constructed using essentially the same upscaling procedure applied by \citep{aliyev2015multilevel}, though here we use the recent implementation of Crain~\citep{crain2020multifidelity}. The NPVs computed from the LF models are corrected using tree-based error models -- specifically random forest (RF) and light gradient boosting machine (LGBM). The RF procedure is similar to that developed by Trehan and Durlofsky \citep{trehan2018machine} within the context of uncertainty quantification, while the LGBM is new in this setting. Our approach shares some similarities with that of Nwachukwu
et al.~\citep{nwachukwu2018fast}, who also used a tree-based method (GB) for WPO. Their treatment, however, did not employ LF models, so it would be expected to require more extensive training for large 3D models (only 2D models were considered in their study). Our method differs from the CNN-based procedures of Kim et al.~\citep{kim2020robust} and Wang et al.~\citep{wang2021efficient} in that we again require substantially less training. Note, however, that our method is much slower during runtime than all of these methods, since we still must perform a large number of flow simulations (because these are all at low fidelity, the computational demands remain reasonable). Finally, because our approach has a clear physics-based component, results may be more explainable than those from purely data-driven methods.

This paper proceeds as follows. In Section~\ref{sec:problem}, we define the well placement optimization (WPO) problem and discuss the differential evolution (DE) optimizer applied in this study. In Section~\ref{sec:LF_model}, the upscaling procedure used to construct the LF models is described. The tree-based error models, and the static and dynamic features considered in these models, are presented in detail. The full offline (preprocessing) and online (runtime) workflows are provided in Section~\ref{sec:strategy}. Optimization results demonstrating the performance of our surrogate treatments are presented in Section~\ref{sec:results}. Two cases are considered, involving different HF geomodels (both are 3D and contain $60 \times 60 \times 30$ cells) and different numbers of injection and production wells to be optimally placed. We conclude in Section~\ref{sec:summary} with a summary and suggestions for future work in this area.

\section{Problem statement and optimization algorithm}
\label{sec:problem}

In this section we present the optimization problem and describe the differential evolution optimizer used in this study. In subsequent sections, the low-fidelity (upscaled) simulation model and the tree-based machine-learning error correction procedure are explained. The overall workflow, which involves all of these components, is then presented.

\subsection{Problem formulation}
\label{sec:formulation}

The well placement optimization (WPO) problem considered here entails the determination of the locations for a specified number of vertical injection and production wells such that a prescribed objective function is maximized or minimized. In this work we seek to maximize the net present value (NPV) of the project over a prescribed production time frame. The operational settings (injection/production rates or wellbore pressures) of the wells are specified, i.e., they are not part of the optimization problem. The formulation here thus represents a reduced version of the general field development problem.

The optimization problem can be expressed as
\begin{align}
\begin{cases}
\underset{\mathbf{u}\in \mathbb{U}}{\max} \;\; \mathit{J}(\mathbf{u}),  \\
\mathbf{c}(\mathbf{u}) \leq \mathbf{0},  \label{eqn:wpo_def}
\end{cases}
\end{align}
where $\mathit{J}$ is the objective function (NPV) to be maximized, $\mathbf{u}\in \mathbb{U} \subset{\mathbb{R}^{2N_w}}$ is the vector containing the $x$--$y$ locations of all $N_w$ wells, and $\mathbb{U}$ defines the feasible regions for well locations. The vector $\mathbf{c}$ denotes any nonlinear constraints. In this study, the only such constraint is a minimum well-to-well distance. The optimization variables include only the $x$--$y$ well locations because all wells penetrate the full model in the $z$-direction. The number of wells ($N_w$) is user-specified.

Following \citep{nasir2021two}, we compute NPV as follows
\begin{equation}\label{eq:npv_def}
    {\text {NPV}} = \sum_{k=1}^{N_t} \frac{\Delta t_k \left[ \sum\limits_{i=1}^{N_{prod}}(p_o q_{o,k}^i - c_{pw} q_{pw,k}^i) - \sum\limits_{i=1}^{N_{inj}} c_{iw} q_{iw,k}^i \right]}{(1+b)^{t_k/365}}.
\end{equation}
Here $N_t$ is the number of time steps in the simulation, $t_k$ and $\Delta t_k$ are the time and time step size (both in days) for step $k$, $N_{prod}$ and $N_{inj}$ are the number of production and injection wells ($N_w=N_{prod}+N_{inj}$), and $q_{o,k}^i$, $q_{pw,k}^i$, and $q_{iw,k}^i$ are the oil/water production and water injection rates of well~$i$ at time step~$k$, respectively. In addition, $p_o$ is the oil price (here specified as 60~USD/bbl), $c_{pw}$ is the cost of produced water (3~USD/bbl), $c_{iw}$ is the cost of injected water (2~USD/bbl), and $b$ is the annual discount rate (here $b=0.1$). 

Because we specify the number of fully penetrating injection and production wells, the well costs are the same in all cases and do not need to be included in the NPV calculation. This is in contrast to the setup in \citep{nasir2021two}. If so desired, the NPVs reported in Section~\ref{sec:results} could be adjusted to include the well costs, though the optimal well locations are independent of the values used.

\subsection{Differential evolution optimization algorithm}
\label{sec:DE}

The differential evolution (DE) procedure used in this work is described by \citep{price2006differential}. DE is a population-based stochastic search procedure with some features that are analogous to those in genetic algorithms (GAs). GAs have been used in a number of earlier WPO studies, as discussed in \citep{nasrabadi2012well} and \citep{onwunalu2010application}. In contrast to some GA implementations, DE uses real numbers rather than binary representations for the optimization variables. Both methods apply operations that are referred to as mutation, crossover, and selection, though the detailed treatments for these procedures differ between DE and GA. We apply DE here rather than GA or particle swarm optimization (PSO) based on the findings of Zou et al.~\citep{zou2021}, who observed DE to provide NPVs comparable to those from PSO, but with less computational effort.

The potential solutions (well configurations) that comprise the population, at DE iteration $k$, are denoted by $\mathbf{u}_i^k \in \mathbb{R}^{2N_w}$, $i =1,\dots,N_p$, where $N_p$ is the size of the population. The set containing all $N_p$ of these solutions is indicated by $P_u^k$. In DE we also have a so-called trial population, $\hat{\mathbf u}_i^k \in \mathbb{R}^{2N_w}$, $i =1,\dots,N_p$. The full set of trial solutions is denoted by ${\hat P}_u^k$.

The solutions in $P_u^k$ are inherited from the previous DE iteration. The trial population ${\hat P}_u^k$ is constructed through application of mutation and crossover operations. Here we briefly explain these procedures. See \citep{price2006differential} or \citep{zou2021} for more details.

Mutation in our DE algorithm (which differs from mutation in GA) can be described by
\begin{align}
\mathbf{x}_i^k = \mathbf{u}_{r,0}^k + s_f\cdot{(\mathbf{u}_{r,1}^k -\mathbf{u}_{r,2}^k)}, \ \ i=1,\dots,N_p. \label{eqn:de_mutation}
\end{align}
Here, $s_f$ is the scale factor, which controls the evolution rate, and $\mathbf{u}_{r,0}^k$, $\mathbf{u}_{r,1}^k$, and $\mathbf{u}_{r,2}^k$ are randomly selected candidate solutions from the population $P_u^k$ (the three $\mathbf{u}_{r}^k$ are all distinct). The $\mathbf{x}_{i}^k$, $i=1,\dots,N_p$, are referred to as the mutant vectors.

A crossover procedure is then applied. This is expressed, for $i=1,\dots,N_p$, $j=1,\dots,2N_w$, as
\begin{align}
(\hat u_{i}^k)_j = 
\begin{cases}
 (x_{i}^k)_j \quad \textrm{if } r_j \leq C_r \\
 (u_{i}^k)_j \quad \textrm{otherwise}. \label{eqn:de_def_trial_pop}
\end{cases}
\end{align}
Here $(\hat u_{i}^k)_j$ indicates a component of $\hat{\mathbf u}_i^k$ and $C_r$ is the user-defined crossover probability. This operation, in which $\hat{\mathbf u}_i^k$ inherits the $\mathbf{x}^k_i$ component each time $r_j \leq C_r$, is referred to as binomial crossover. The result of this process is the population of trial solutions, ${\hat P}_u^k$. 

Now, given the full sets of (corresponding) inherited and trial solutions, $\mathbf{u}_{i}^k$ and $\hat{\mathbf u}_i^k$, $i=1,\dots,N_p$, we select the solution to advance to $P_u^{k+1}$, the population at the next iteration. This selection is determined simply by choosing the solution that provides the higher NPV. This requires us to perform flow simulations for all of the $N_p$ trial solutions in ${\hat P}_u^k$. This is by far the most expensive step in the overall procedure. We do not need to simulate the $N_p$ solutions in $P_u^k$ because these were evaluated at the previous iteration. The sequence of treatments thus described (mutation, crossover, selection) is continued until a prescribed termination criterion, such as maximum number of DE iterations or minimum improvement in NPV over some number of iterations, is met.

The total number of flow simulations required during DE optimization is equal to $N_p \times N_{iter}$, where $N_{iter}$ is the number of iterations performed. If all of these simulations are conducted using HF models, the cost of the optimization can be excessive, as each run may require hours in practical cases. In the next section, we will describe surrogate and error-modeling treatments that substantially reduce the computations required for function evaluations.
\section{Low-fidelity surrogate with error correction}
\label{sec:LF_model}

In this section, we first describe the construction of the LF (upscaled) flow simulation models. The random forest (RF) and light-gradient boosting machine (LGBM) error-correction treatments, applied to improve the LF function evaluations, are then discussed. Next, the detailed features used in the machine-learning algorithms are presented.

\subsection{Construction of LF models via upscaling}
\label{sec:upsc}

The computational time required for each flow simulation (needed to evaluate NPV in Eq.~\ref{eq:npv_def}) is a function of the number of grid blocks/cells in the simulation model. Aside from simulator overhead, which represents a fixed computational cost, simulation time scales at least linearly with the number of blocks. This scaling may be super-linear depending on the flow physics, model type, etc. Thus, it is clearly beneficial to reduce the number of cells in the simulation model.

Because model properties can vary strongly and discontinuously from block to block due to geologic heterogeneity, simple coarsening or averaging procedures may provide very inaccurate LF models. We therefore apply global transmissibility upscaling. This procedure requires a HF flow problem to be solved, but this computation is very fast compared to the simulations required to evaluate NPV, as we now explain.

In the problem setup considered in this work, water is injected to produce oil. We denote the number of cells in the HF problem as $N_h$. The two-phase flow simulation (see \citep{trehan2018machine} for the governing equations and a brief discussion of the finite volume discretization procedure) entails the solution of $2N_h$ nonlinear equations at each time step. These equations express oil and water conservation in each grid block, and the associated unknowns are the pressure and water saturation in each block. The nonlinear equations are solved using Newton's method. The computational complexity is therefore $O(2N_hN_tN_n)$, where $N_n$ is the average number of Newton iterations per time step ($N_n$ is typically $\sim$2--5) and $N_t$ is the number of time steps. Note that $N_t$ is commonly $\sim$100.

To construct the LF model, we solve a single-phase HF flow problem at a single time step. This enables us to compute upscaled flow parameters. The computational complexity of this operation is $O(N_h)$, which is much less than that associated with the two-phase HF flow simulations. The computational complexity for two-phase flow simulation with the LF model is $O(2N_lN_tN_n)$, where $N_l$ is the number of cells in the LF model. Because $N_l \ll N_h$, the total computation required to construct and simulate the LF models is very small compared to the requirements for two-phase flow simulation with HF models. We note finally that the LF models still have error associated with them because only single-phase flow parameters are upscaled. Upscaled two-phase flow parameters (e.g., relative permeabilities) could also be computed, and this would provide more accurate LF models. The requisite computations are expensive, however, so we do not proceed in this manner.

We now provide a brief description of the upscaling procedure applied in this work. The specific implementation is that of Crain \citep{crain2020multifidelity}, which is based on approaches developed by \citep{zhang2008new} and \citep{chen2008nonlinear}. Please consult these references for full details.

We first solve the single-phase pressure equation over the full 3D domain, with flow driven by the wells in the configuration under consideration (by this we mean the wells associated with ${\mathbf u}_i^k$ or $\hat{\mathbf u}_i^k$):
\begin{equation}\label{eq:single_phase_Darcy}
    \begin{split}
        \nabla \cdot \left(\frac{\mathbf{k}}{\mu} \nabla {p} \right) = q.
    \end{split}
\end{equation}
Here $\mathbf{k}$ denotes the permeability tensor, here taken to be diagonal, $\mu$ is the fluid viscosity, ${p}$ is pressure, and $q$ is the well-driven source term, which depends on ${\mathbf u}_i^k$ or $\hat{\mathbf u}_i^k$ and the specified well pressures. Note that $\mathbf{k}$ can vary strongly from block to block. The value of $\mu$ does not impact the LF quantities computed below, and we set $\mu=1$ without loss of generality. Eq.~\ref{eq:single_phase_Darcy} is discretized and solved over the HF (fine) grid, which contains $N_{h}$ cells.

Following the solution of the discretized version of Eq.~\ref{eq:single_phase_Darcy}, we compute the LF model properties such that the integrated (summed) fluxes over the HF regions corresponding to each LF interface are retained. This involves the computation of coarse-scale transmissibility, which is essentially the numerical analog of permeability. This calculation can be expressed as
\begin{align}
T^*_j = \frac{\sum_l f_l}{\langle p \rangle_{j} - \langle p \rangle_{j+1}}. \label{eqn:upsc_computation_trans}
\end{align}
Here $T^*_j$ is the upscaled transmissibility linking (adjacent) LF cells $j$ and $j+1$, $\langle p \rangle_{j}$ denotes the pressure averaged over the HF cells that lie within LF cell $j$ (and similarly for $\langle p \rangle_{j+1}$), and $f_l$ denotes the flow rate through HF cell faces $l$ that lie on the interface between LF cells $j$ and $j+1$.

The well index is the property that links the well to the grid blocks it penetrates (these are referred to as wellblocks). This property is analogous to the block-to-block transmissibility explained above. From the discrete solution of Eq.~\ref{eq:single_phase_Darcy}, we compute the LF well indices through application of
\begin{align}
WI^*_j = \frac{\sum_l f_l^w}{\langle p \rangle_{j} - p^w}, \label{eqn:upsc_computation_well_idx}
\end{align}
where $WI^*_j$ is the well index for LF wellblock $j$, $p^w$ is the well pressure evaluated at the vertical midpoint of wellblock $j$, and $f_l^w$ is the flow rate into or out of the well in HF cell $l$. The sum is over all HF cells that fall within the LF wellblock.

\begin{figure*}[!htb]
\centering
\begin{subfigure}{.47\linewidth}\centering
\includegraphics[width=\linewidth]{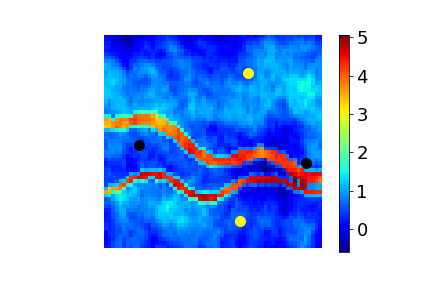}\caption{HF transmissibility map}
\end{subfigure}
\begin{subfigure}{.47\linewidth}\centering
\includegraphics[width=\linewidth]{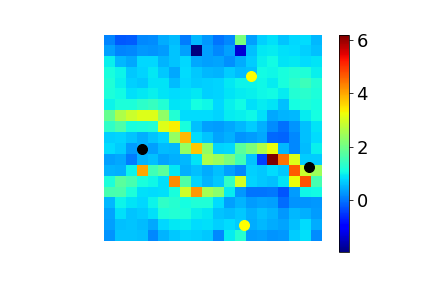}\caption{LF transmissibility map}
\end{subfigure}
\caption{HF and LF transmissibility maps (in units of millidarcy-ft) for a single $x$--$y$ layer. HF layer contains $60 \times 60$ cells and LF layer contains $20 \times 20$ cells. Injection and production wells are denoted by the black and yellow circles.}
\label{fig:upscale_scheme}
\end{figure*}

The application of this upscaling procedure is illustrated in Fig.~\ref{fig:upscale_scheme}. The model contains four wells, indicated by the yellow and black circles. Here a single ($x$--$y$) layer of the HF model, of dimensions $60 \times 60$, is upscaled to an LF layer of dimensions $20 \times 20$. The overall upscaling factor here is thus 9. Larger upscaling factors will be applied in our examples, as we also coarsen in the $z$-direction. The high-transmissibility channel features evident in the HF model (Fig.~\ref{fig:upscale_scheme}a) are not as sharp in the LF model (Fig.~\ref{fig:upscale_scheme}b). The single-phase flow-rate properties are largely retained, however. The errors that result from the use of LF models (as in Fig.~\ref{fig:upscale_scheme}b) are often associated with two-phase flow phenomena such as the imperfect resolution of fluid fronts. This leads to inaccuracy in, e.g., the breakthrough of injected water at production wells.

It is important to reiterate that, because the properties of the LF model depend on the well locations, the $T^*_j$ and $WI^*_j$ must be computed separately for each well configuration (${\mathbf u}_i^k$ or $\hat{\mathbf u}_i^k$, $i=1, \dots, N_p$, $k=1, \dots, N_{iter}$) evaluated during the DE optimization. Although this is much more involved than a simple analytical averaging procedure, the computational requirements are small compared to the two-phase flow simulations performed to compute NPV (even with LF models). Within the DE optimization framework, well locations are always tracked on the HF model, even when the flow simulations are performed using LF models. Wells are assumed to lie at the centers of HF wellblocks, but they can be off-center in LF wellblocks, as shown in Fig.~\ref{fig:upscale_scheme}b. The effect of this (LF) off-centering is captured in the $T^*_j$ and $WI^*_j$ computations.

\subsection{Tree-based error correction of LF model}
\label{sec:error_mod}

The upscaled (LF) model provides improved computational efficiency relative to the HF model, but the resulting predictions still have some error. To mitigate the impact of this reduction in accuracy, we apply tree-based machine-learning to estimate model error. Following \citep{forrester2009recent}, we write
\begin{equation}\label{eqn:model_error_general}
   J_{h}(\mathbf{u}) = J_{l}(\mathbf{u}) + e(\mathbf{u}),
\end{equation}
where $\mathbf{u}$ represents the design variables (well locations), $J_{h}$ and $J_{l}$ denote the objective function values for the HF and LF models, and $e$ is the model error that results from performing function evaluations with the LF model. Our goal is to construct tree-based methods to estimate $e(\mathbf{u})$.

Tree-based techniques are machine-learning (ML) procedures applicable for both regression and classification. Tree-based ML models can be built with much less data than is required for deep neural networks, though they can still represent complicated relationships that could be challenging to capture with traditional regression methods. Thus these approaches are well suited for the WPO problem under consideration, as we seek to train these models with $O(100)$ samples, rather than the many thousands often needed for deep neural networks.

The random forest (RF) procedure used in this study shares similarities with the RF model developed by Trehan and Durlofsky \citep{trehan2018machine}. Both treatments are applicable for the representation of error in LF reservoir simulation predictions, and there is some overlap in the features considered. There are also important differences. Most notably, here we consider varying well locations with a fixed permeability field, while in \citep{trehan2018machine} the well locations were fixed but the permeability fields were varied (to account for geological uncertainty). In addition, the goal in the previous study was to correct the detailed time-varying well-by-well production and injection responses, while here we focus on modeling the error in a single scalar quantity, NPV, and on using the corrected NPV for optimization. Finally, we consider light-gradient boosting machine (LGBM) methods in addition to RF, while the earlier investigation used only RF.

Tree-based models are composed of a set of decision trees. Each tree contains nodes, which represent decision points, and branches (paths), which connect the nodes. A group of nodes at the same level within the tree is referred to as a layer. More nodes and layers are required as the complexity of the problem increases. The input to the trees is a set of features. In an efficient implementation, these are readily available or easily computable quantities that impact the output quantity of interest (QoI). Features with more impact on QoI are more likely to be associated with nodes toward the top of the tree (upper layers), while those that are less important tend to appear toward the bottom.

The two common ways of organizing decision trees are referred to as bagging and boosting. Bagging organizes multiple decision trees in parallel, with the final prediction computed as the average of the predictions from each tree. Boosting, by contrast, treats multiple decision trees sequentially, with each tree predicting the residual (error) from the result of the previous tree. In this work, we apply widely used methods of each type, namely RF (bagging) and LGBM (boosting). For RF we use the implementation of \citep{pedregosa2011scikit} (scikit-learn library), while for LGBM we use that by \citep{ke2017lightgbm} (LightGBM library).

\begin{figure}[!htb]
\centering
\includegraphics[width = 0.48\textwidth]{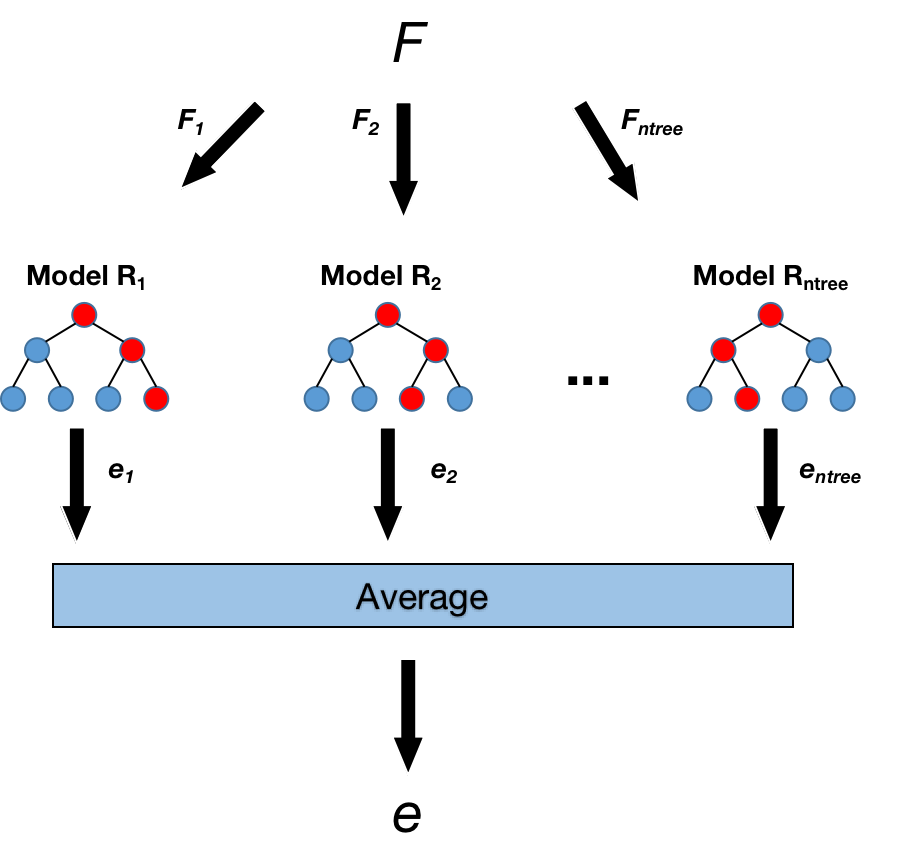}
\caption{Schematic of RF prediction procedure.}\label{fig:random_forest}
\end{figure}

Prediction using RF is illustrated in Fig.~\ref{fig:random_forest}. Here $F$ is the full set of input features, which will be described in Section~\ref{sec:features}. Subsets of $F$, denoted $F_i \subset F$, $i=1, \dots n_{tree}$, are input to each of the $n_{tree}$ trees. The use of different $F_i$ decreases redundancy (and correlation) among the trees and prevents a few features from being overly dominant. Each tree provides a prediction for error $e_i$ (for use in Eq.~\ref{eqn:model_error_general}), and the average over all trees provides the RF estimate. The red points on each tree in Fig.~\ref{fig:random_forest} indicate the path for that particular tree. The different paths for the various trees reflect the fact that different features are evaluated at each node.

\begin{figure}[!htb]
\centering
\includegraphics[width = 0.48\textwidth]{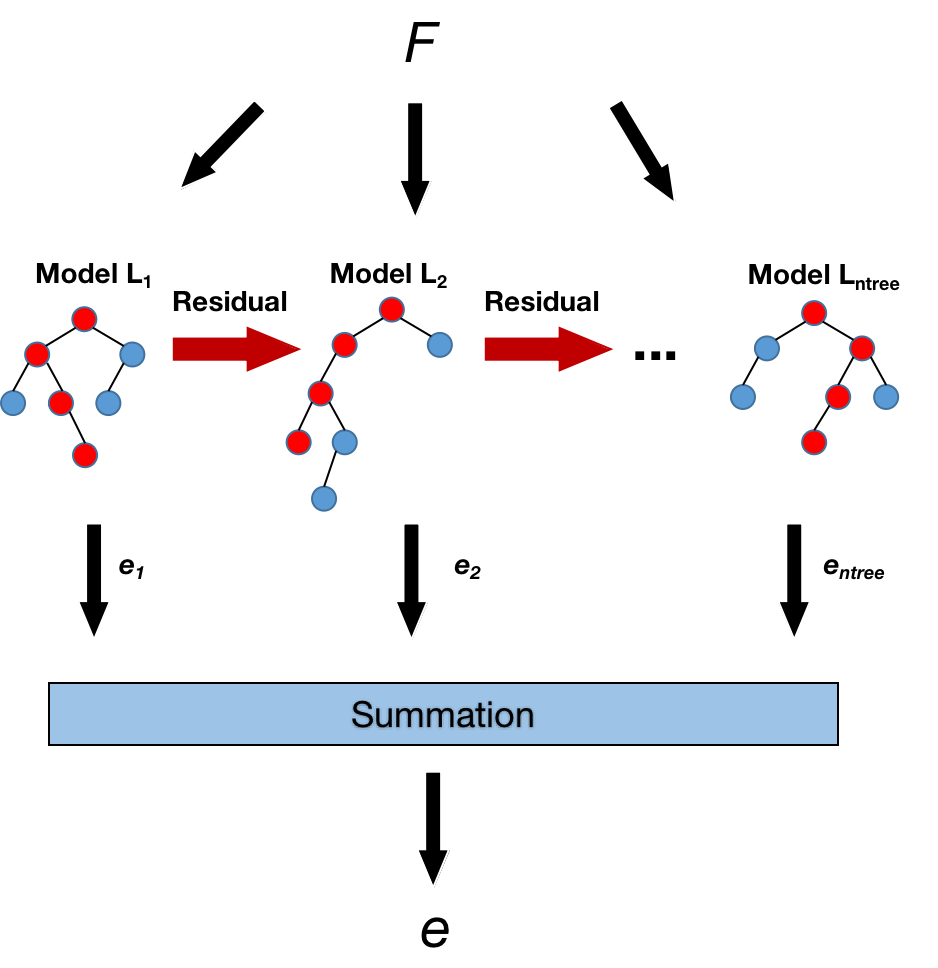}
\caption{Schematic of LGBM prediction procedure.} \label{fig:gradient_boosting}
\end{figure}

Prediction based on boosting is illustrated in Fig.~\ref{fig:gradient_boosting}. Rather than averaging the predicted results from each tree as in bagging, in boosting we sum the results from each tree, as each tree here is trained to correct the residual from the previous tree. As noted by \citep{ke2017lightgbm}, LGBM has been shown to outperform many other boosting methods. With LGBM, instead of building each tree by completing all previous layers, the tree is constructed in the direction of maximum reduction of prediction error. This is referred to as leaf-wise tree splitting, in contrast to level-wise tree splitting in RF (Fig.~\ref{fig:random_forest}). Here a leaf is defined as the end node of each tree model, e.g., the bottom-most red nodes in Fig.~\ref{fig:random_forest}. 

To be more specific, tree ${\text R}_1$ in Fig.~\ref{fig:random_forest} contains 7~nodes and 4~leaves, and the depth of this tree is 3~layers. Successive layers are only constructed after the previous layer is full. Tree ${\text L}_2$ in Fig.~\ref{fig:gradient_boosting} also contains 7~nodes, but it has 3~leaves and the tree depth is 5~layers. This results from building the tree in the direction of fastest reduction of the prediction error. The leaf-wise splitting of trees can lead to an increase in model flexibility and performance. This strategy may, however, lead to over-fitting, though this can be mitigated through careful hyperparameter tuning. 

Although RF and LGBM represent popular bagging and boosting methods, it is not always clear which is preferable in a particular setting. We found that, for our application, performance was enhanced by averaging the predictions from the two methods. More specifically, for a large set of test cases, we observed that this approach led to less average error and less variance in error than using either method by itself. Therefore, we use the average of the predictions from both methods for our estimate of $e$ in Eq.~\ref{eqn:model_error_general}.

\subsection{Error model features and training}
\label{sec:features}

Before describing the features used in the tree-based models, we explain how the problem is decomposed. From detailed numerical tests, we found that improved results could be achieved by computing corrections on a well-by-well basis rather than constructing a single global correction to NPV. This means that we write ${\text{NPV}}=\sum_{i=1}^{N_w} {\text{NPV}}_i$, where ${\text{NPV}}_i$ denotes the contribution associated with each well. This is computed as in Eq.~\ref{eq:npv_def}, but without the sum over $N_{prod}$ or $N_{inj}$. We then construct $N_w$ separate ML models to correct each of these ${\text{NPV}}_i$. All but one of the features associated with each model are local to that particular well.

Feature selection is a very important aspect of any ML method. Here we follow a strategy similar to that described in \citep{trehan2018machine} for feature selection. We tested a range of features, and found the set of features listed in Table~\ref{tab:features_table} to be the most relevant. Note that some features used in purely data-driven surrogates (e.g., \citep{nwachukwu2018fast}), such as well-to-well distances, are not used in our framework. This is presumably because these features are less informative when the goal is to correct a LF result, as it is in our case.

We now explain the features appearing in Table~\ref{tab:features_table}. Feature~1 is based on the input geomodel. The next set of features (2, 3, 4 and 5) are computed when we solve the single-phase steady-state problem used for upscaling.  The remaining features derive from the coarse-scale two-phase flow simulation. Feature~1, $\mathbf{K} \in {\mathbb{R}^{N_{wb}^h}}$, represents the value of permeability at all well blocks in the fine-scale geomodel (the total number of wellblocks in the HF model is denoted $N_{wb}^h$). Note that the dimensions given in this discussion are for the features corresponding to all wells -- for a single well, there are $N_z$ elements in $\mathbf{K}$, where $N_z$ is the number of layers in the model. The vector $\mathbf{K}$ contains one element per wellblock because permeability is here taken to be of the form $k_x=k_y$, $k_z= \gamma k_z$, where $k_x$, $k_y$, and $k_z$ denote permeability components in the coordinate directions, and $\gamma$ is a fixed constant.

The next set of features, $\mathbf{T}^* \in {\mathbb{R}^{N_{tr}}}$ and  $\mathbf{WI}^* \in {\mathbb{R}^{N_{wb}^l}}$, are the outputs from the upscaling procedure. Here $N_{tr}$ is the total number of cells adjacent to the wellblocks and $N_{wb}^l$ is the number of wellblocks in the LF model. If all wells are at least one coarse block away from boundaries, we have $N_{tr}=6N_z-2$ (with the $-2$ appearing because the top and bottom layers have only one neighbor in $z$). Next, we have two features based on flow quantities computed from the HF single-phase flow pressure solution -- $\mathbf{V}_{1p}\in {\mathbb{R}^{3N_{wb}^h}}$ and $\mathbf{V}^\prime_{1p}\in {\mathbb{R}^{3N_{wb}^h}}$. The vector $\mathbf{V}_{1p}$ contains the $x$, $y$, and $z$ components of velocity for each HF well block, while $\mathbf{V}^\prime_{1p}$ contains the `fluctuating' HF wellblock velocity components (these fluctuating components were also found to be informative in \citep{trehan2018machine}). The components of $\mathbf{V}^\prime_{1p}$ are computed as the difference between the HF wellblock velocity and the average velocity over all HF wellblocks that lie within a LF wellblock. 

The last set of features derive from the LF two-phase flow solution. These comprise the total velocity in the wellblocks at each time step, $\mathbf{V}^{l} \in {\mathbb{R}^{3N_{wb}^l \times N_{t}}}$, where total velocity is the sum of the oil-phase and water-phase velocities, and the wellblock pressure and water saturation at each time step, $\mathbf{P}^{l}\in {\mathbb{R}^{N_{wb}^l \times N_{t}}}$ and $\mathbf{S}_w^{l}\in {\mathbb{R}^{N_{wb}^l \times N_{t}}}$. The final feature is pore volume injected, $\mathbf{PVI}^{l} \in {\mathbb{R}^{N_{t}}}$, computed up to the time corresponding to each time step. PVI is essentially a dimensionless time and is computed via ${\text{PVI}}^l_k= {\frac{1}{V_p}} \left(\int_0^{t_k} q dt \right)$, $k=1,\dots,N_t$, where $q$ is the field water injection rate (summed over all $N_{inj}$ injection wells) and $V_p$ is the total pore volume of the reservoir. Note that PVI is the only global feature shared in all tree-based models.

The training procedure used in this work is as described by \citep{rokach2007data} and \citep{mantovani2018empirical}. Tuning parameters include the learning rate, total number of trees, maximum depth of any tree, and the number of leaves in each tree. In addition, because we use only $O(100)$ training samples, we need to limit the number of features considered to avoid over-fitting. This is accomplished through application of principal component analysis (PCA) to the features involving time-series ($\mathbf{V}^{l}$, $\mathbf{P}^{l}$, $\mathbf{S}_w^{l}$ and $\mathbf{PVI}^{l}$). PCA is applied to each such feature individually. The number of PCA components retained is also treated as a tuning parameter during the training process. The parameters tuned during training are varied using a grid search strategy. Training continues until a prescribed error tolerance is reached. 

Finally, we note that features~1--8, currently evaluated only at wellblocks, could be extended to include additional variables at one or more `rings' around the wellblocks. This could lead to improved performance as these near-well quantities are also expected to impact the error associated with LF simulation. In this case, PCA should be applied to these sets to limit the total number of features considered by RF and LGBM.

\begin{table}[htb!]
\centering
    \caption{Features used in the error models}
    \begin{tabular}{|c|c|c|}
    \hline
    {No.} & {Feature} & {Quantity and Fidelity Level} \\ \hline
    {1.} & {$\mathbf{K}$} & {permeability (HF)}\\ 
    {2.} & {$\mathbf{T}^*$} & {transmissibility (LF)}\\ 
    {3.} & {$\mathbf{WI}^*$} & {well index (LF)}\\ 
    {4.} & {$\mathbf{V}_{1p}$} & {single-phase steady-state velocity (HF)}\\
    {5.}  & {$\mathbf{V}^\prime_{1p}$} & {single-phase velocity fluctuation (HF)  }\\
    {6.} & {$\mathbf{V}^{l}$} & {total velocity (LF)}\\ 
    {7.} & {$\mathbf{S}_w^{l}$} & {water saturation (LF)}\\ 
    {8.}  & {$\mathbf{P}^{l}$} & {pressure (LF)} \\
    {9.} & {$\mathbf{PVI}^{l}$} & {pore volume injected (LF)}\\ 
    \hline
    \end{tabular}%
    \label{tab:features_table}%
\end{table}%

\section{Optimization using LF model and ML error estimation}
\label{sec:strategy}
In this section, we describe the detailed offline (training) and online (runtime) workflows used in our implementation. These are also provided in  Algorithm~\ref{alg:wpo_strategy}. The goal of our framework is to perform WPO with LF models, but to correct the NPVs computed from LF results using the tree-based ML methods described in Section~\ref{sec:error_mod}. We use `LF+corr' to refer to the LF model result plus the error correction. 

\begin{algorithm*}[!htpb]{
\caption{Well placement optimization using the LF+corr model} \label{alg:wpo_strategy}
\SetAlgoVlined
\textbf{Offline Stage:} Build the error model \\
\For{$j \leq N_{opt}^{0}$} 
        {Start DE run $j$ \\
        Initialize population $P_u^0$ with random seed $j$; \\
        \While{$I_n^l > \epsilon$}
        {\For{$\mathbf{u}_i^k \in P_u^k$} 
        {Perform mutation (Eq.~\ref{eqn:de_mutation}) and crossover (Eq.~\ref{eqn:de_def_trial_pop}) to obtain $\hat{\mathbf u}_i^k$ from $\mathbf{u}_i^k$; \\
        Apply upscaling to construct the LF model with $\hat{\mathbf u}_i^k$; \\
        Simulate the LF model and compute ${\text {NPV}}_l$;\\
        Update $\mathbf{u}_i^{k+1}$ from $\mathbf{u}_i^k$ and $\hat{\mathbf u}_i^k$ based on ${\text {NPV}}_l$; \\
        }
        Collect well scenarios from $P_u^k$;\\
        Assemble $\mathbf{u}_i^{k+1}$, $i=1,\dots,N_p$, to give $P_u^{k+1}$ 
        }
        }
Cluster all well scenarios considered into $N_{cl}$ clusters;\\
Select $N_{cl}$ well scenarios as training samples from clusters;\\
Train RF and LGBM error models with $N_{cl}$ well scenarios;\\
\textbf{Online Stage:} Use the LF+corr surrogate for WPO\\
\For{$j \leq N_{opt}$} 
        {Start DE run $j$\\
        Initialize the population $P_u^0$ with random seed $j$; \\
        \While{$I_n^l > \epsilon$}
        {\For{$\mathbf{u}_i^k \in P_u^k$} 
        {Perform mutation (Eq.~\ref{eqn:de_mutation}) and crossover (Eq.~\ref{eqn:de_def_trial_pop}) to obtain $\hat{\mathbf u}_i^k$ from $\mathbf{u}_i^k$; \\
        Apply upscaling to construct the LF model with $\hat{\mathbf u}_i^k$; \\
        Simulate the LF model and compute ${\text {NPV}}_l$;\\
        Apply tree-based error models to estimate error $e$;\\
        Use LF+corr result (${\text {NPV}}_l+e$) as NPV estimate;\\
        Update $\mathbf{u}_i^{k+1}$ from $\mathbf{u}_i^k$ and $\hat{\mathbf u}_i^k$ based on NPV estimates; \\
        }
        Assemble $\mathbf{u}_i^{k+1}$, $i=1,\dots,N_p$, to give $P_u^{k+1}$
        }
        }
Save $\mathbf{u}^{opt}$ from the online optimization. Perform HF simulation with $\mathbf{u}^{opt}$ to obtain $\text{NPV}_h^{opt}$. \\
}
\end{algorithm*}

ML-based models often display reduced accuracy outside of their training range. In optimization problems, however, the goal is to evolve the solution towards more promising portions of the search space, which often correspond to regions away from the training range. Several strategies can be used to address this issue. An obvious choice is to occasionally retrain the ML model, as was done in \citep{trehan2016trajectory} for well control optimization problems. This approach is effective but it leads to additional computation as some number of new HF simulations, followed by retraining, are required. Alternatively, when the ML model loses accuracy, the algorithm can simply shift to HF flow simulation. This approach, which was used in \citep{kim2021recurrent} for well control optimization, can become expensive if a large number of HF flow simulations are required.

We tested the two strategies described above along with a third approach. This third approach, which performed the best and is used in this work, is as follows. As part of the offline procedure, we first perform a substantial number of LF optimization runs (here we conduct $N_{opt}^{0}=25$ of these runs, each with a different randomly generated initial population). These optimizations do not include any error correction, so the resulting NPVs are somewhat inaccurate, but the well configurations provided in the later iterations of these runs correspond to reasonable solutions. A clustering procedure is then applied to select a subset of the configurations generated during these runs. HF simulations with these configurations are then performed. The set of corresponding HF and LF models and solutions comprise the training samples used to train the RF and LGBM models. During training (discussed in Section~\ref{sec:features}), parameters are varied to reduce the average model error $e(\mathbf{u})$ to a specified tolerance $e_{tol}$.

Our clustering procedure is as follows. During the offline stage, we collect $N_{opt}^{0}\times N_{eval}$ solutions ($\sim$75,000 well configurations for the current study), where $N_{eval}$ is the number of function evaluations for a single optimization run. Note that $N_{eval}$ can vary based on when the optimization run terminates, though here it is typically $\sim$3000. The number of clusters ($N_{cl}$) is specified to be equal to the number of training samples, which is either 100 or 150 in the cases considered in Section~\ref{sec:results}. We apply k-means clustering with the full set of $\mathbf{u}_i$, $i=1,\dots,N_{opt}^{0}\times N_{eval}$. Once $N_{cl}$ clusters are formed, the three $\mathbf{u}_i$ in each cluster that correspond to the highest NPVs (based on LF simulation results) are identified. For each cluster, we then select the $\mathbf{u}_i$ that is furthest (in terms of L$_2$ distance) from all of the previously selected $\mathbf{u}_i$. With this approach, we identify $N_{cl}$ training samples that are diverse in terms of well configurations and that correspond to relatively high NPVs. This is beneficial as high NPV solutions are most relevant to the online optimization procedure.

The optimization termination criterion used in this study, for both the offline and online optimizations, follows the approach of \citep{de2021field}. Specifically, we define the relative objective function improvement, $I_n^l$, as
\begin{equation}\label{eqn:opt_termination_criteria}
    I_{n}^l = \frac{J_{n}^l-J_{n}^{l-1}}{J_{n}^{l-1}}, \ \ l = 2,\dots,l_{max},
\end{equation}
where $n$ is the user-defined interval, in terms of number of function evaluations, after which the improvement $I_n^l$ is evaluated, $J_{n}^l$ denotes the best objective function value obtained thus far (after $l \times n$ function evaluations have been performed), $J_{n}^{l-1}$ is the best objective function value obtained at the end of the previous interval (after $(l-1) \times n$ function evaluations), and $l_{max}$ is chosen such that a specified maximum number of function evaluations is not exceeded. Whenever $n$ function evaluations have been completed, we calculate $I_n^l$ and terminate the optimization if $I_n^l \leq \epsilon$, where $\epsilon$ is a specified tolerance. If $I_n^l > \epsilon$, we proceed with the optimization, unless a maximum number of function evaluations has been reached.

In Algorithm~\ref{alg:wpo_strategy}, ${\text {NPV}}_l$ denotes the NPV computed from simulation on a LF model, $\mathbf{u}^{opt}$ is the optimum configuration as determined from the online (LF+corr) optimization, and $\text{NPV}_h^{opt}$ is the NPV computed from simulation of configuration $\mathbf{u}^{opt}$ on the HF model. In addition, $N_{opt}$ is the number of online optimization runs performed, which can differ from $N_{opt}^0$. Note that in Section~\ref{sec:results} we perform HF simulation on the $\mathbf{u}^{opt}$ configurations from each of the $N_{opt}$ optimization runs.
\section{Optimization results}
\label{sec:results}

\begin{figure*}
\begin{minipage}{.32\textwidth}
  \centering
  \includegraphics[width=1.2\linewidth]{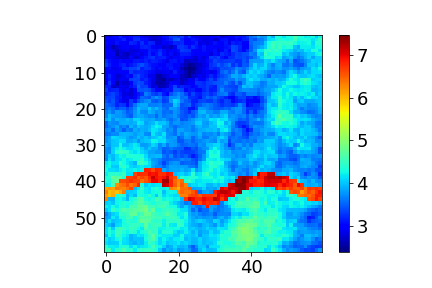}  
  \subcaption{Layer~5}
  \label{fig:res_model_4w_layer5}
\end{minipage}
\begin{minipage}{.32\textwidth}
  \centering
  \includegraphics[width=1.2\linewidth]{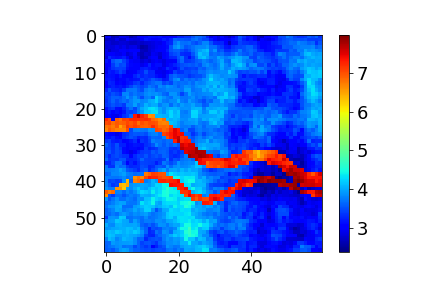}
  \subcaption{Layer~10}
  \label{fig:res_model_4w_layer10}
\end{minipage}
\begin{minipage}{.32\textwidth}
  \centering
  \includegraphics[width=1.2\linewidth]{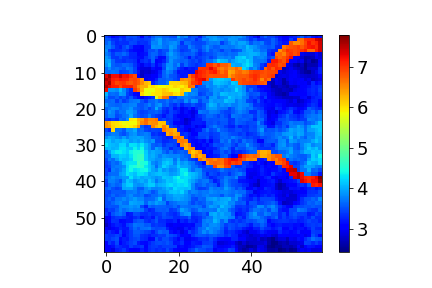}
  \subcaption{Layer~15}
  \label{fig:res_model_4w_layer15}
\end{minipage}
\caption{Permeability field ($\log_e k_x$, with $k_x$ in millidarcy) for three layers of the HF geomodel (Case~1).}
\label{fig:res_model_4w}
\end{figure*}

In this section, we present optimization results for two example cases. The two cases involve different geological realizations (drawn from the same geological scenario) and different numbers of wells. Results will be presented using HF models (we refer to this as HF optimization), using LF models (referred to as LF optimization), and with the LF+corr procedure. In both cases, the 3D (fine-scale) HF reservoir model contains $60 \times 60 \times 30$ grid cells, for a total of 108,000 blocks. The dimensions of each HF cell are $\Delta x = \Delta y = 20$~m and $\Delta z = 5$~m. The HF geomodels represent a reservoir containing channel features, with permeability displaying (different) normal distributions in the high-permeability sand (channel) and low-permeability mud (background) rock types. We set $k_x=k_y$ and $k_z= 0.1 k_x$. The (upscaled) LF models, generated using the implementation of \citep{crain2020multifidelity}, contain $10 \times 10 \times 10$ cells. Each LF block is thus of dimensions $\Delta x = \Delta y = 120$~m and $\Delta z = 15$~m. Porosity is 0.3 in all cells. All flow simulations are performed using an internally developed simulator, ResSimAD, written by Yimin Liu. The optimizations are conducted within the Stanford Unified Optimization Framework.

\begin{figure*}[!htpb]
\begin{subfigure}{.5\textwidth}
  \centering
  \includegraphics[width=0.95\linewidth]{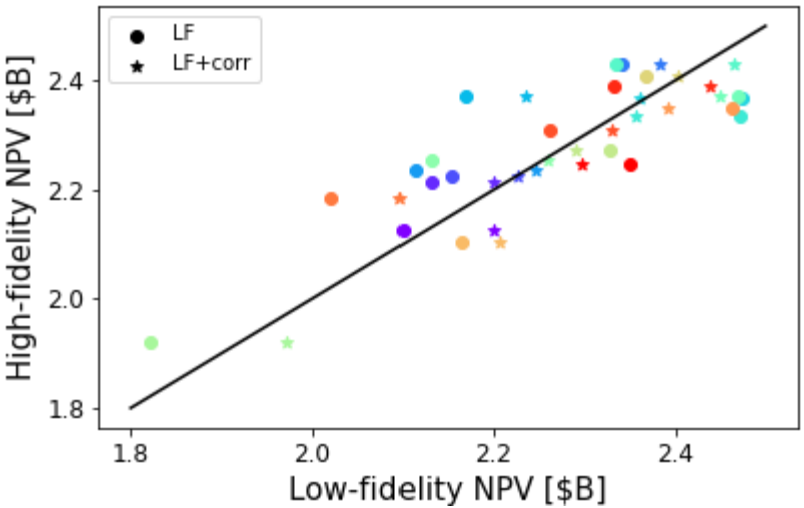}
  \caption{Cross-plots of NPV evaluated on HF, LF, and LF+corr models}
  \label{fig:error_model_4w_dot}
\end{subfigure}
\begin{subfigure}{.5\textwidth}
  \centering
  \includegraphics[width=0.95\linewidth]{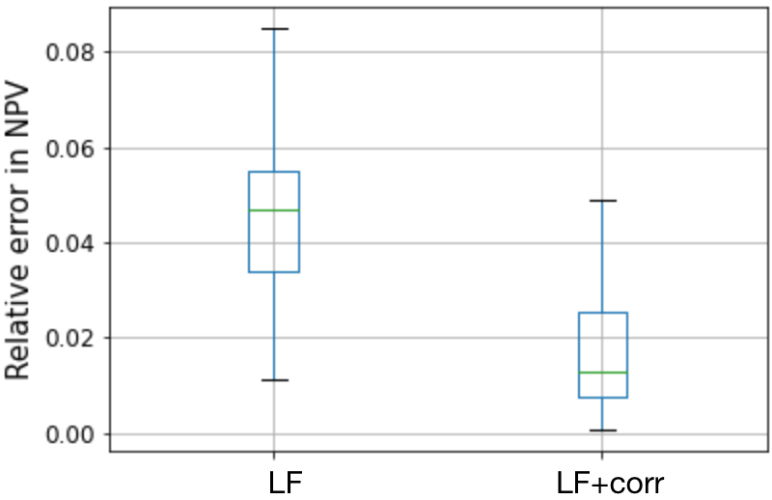}
  \caption{Box plots of relative error}
  \label{fig:error_model_4w_box}
\end{subfigure}
\caption{Test sample results for Case~1.}
\label{fig:error_model_4w}
\end{figure*}

\subsection{Case~1: two injectors and two producers}
\label{sec:case1}

We simulate oil-water flow, with an initial water saturation of 0.1 throughout the model. The oil phase contains a substantial amount of dissolved gas, and the overall system is compressible. The initial pressure in the top layer of the model is $200$~bar. Initial pressure in lower layers varies with depth. The bottom-hole pressure (BHP) of the injection wells is set to $250$~bar, while BHP for production wells is specified as $150$~bar. Note BHP corresponds to the wellbore pressure at the top of the formation. The simulator models the variation in wellbore pressure with depth based on the density of fluid in the wellbore.

\begin{figure}[!htpb]
\centering
\includegraphics[width = 0.48\textwidth]{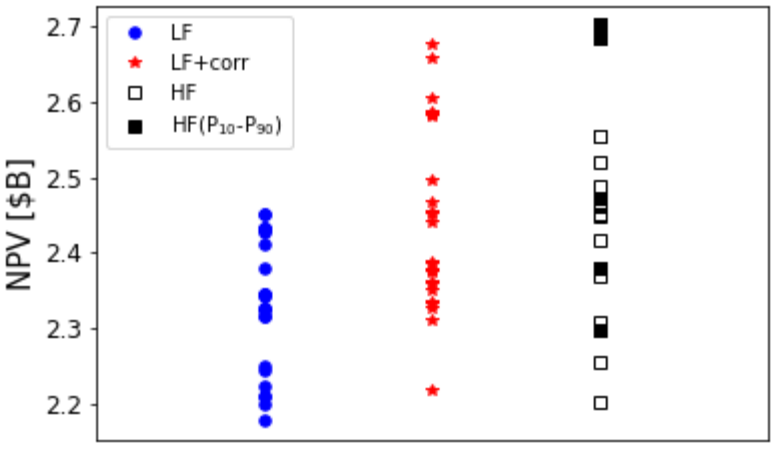}
\caption{Optimization results for Case~1. All NPVs are evaluated on the HF model.} \label{fig:wpo_4w_final}
\end{figure}

In the offline stage of Algorithm~\ref{alg:wpo_strategy}, we perform $N_{opt}^0 = 25$ optimization runs with LF models. Different randomly generated initial populations are used in each run. In both the offline and online stages, following guidelines in \citep{zou2021}, the population size is set to $N_p = 24$, the scale factor to $s_f=0.5$, and the crossover probability to $C_r = 0.7$. The optimization is terminated when the change in NPV is less than $\epsilon = 1\%$ over 60 iterations (offline stage) or 80 iterations (online stage).

The permeability fields for three layers of the geomodel used in Case~1 are shown in Fig.~\ref{fig:res_model_4w}. In this case, we specify $N_w = 4$~wells (two injectors and two producers). The total simulation time frame is $3000$~days.

The error model is constructed using 100 different well scenarios considered during the offline LF optimization procedure. The $N_{cl}=100$ samples are divided into 80 training samples and 20 test samples. Error model performance for the 20 test samples is shown in Fig.~\ref{fig:error_model_4w}a. Here the $y$-axis value is the NPV for a particular scenario evaluated using the HF model (this is the reference result), and the $x$-axis value is the NPV for the corresponding well scenario evaluated using either the LF or LF+corr model. The LF result (without correction) for a particular well configuration is indicated by a solid-colored circle, while the corresponding LF+corr result is denoted by a colored star. For a given scenario, we use the same color for the LF and LF+corr results, so the correction for each of the 20 test samples is clearly evident in Fig.~\ref{fig:error_model_4w}a.

\begin{figure}
\centering
\includegraphics[width = 0.48\textwidth]{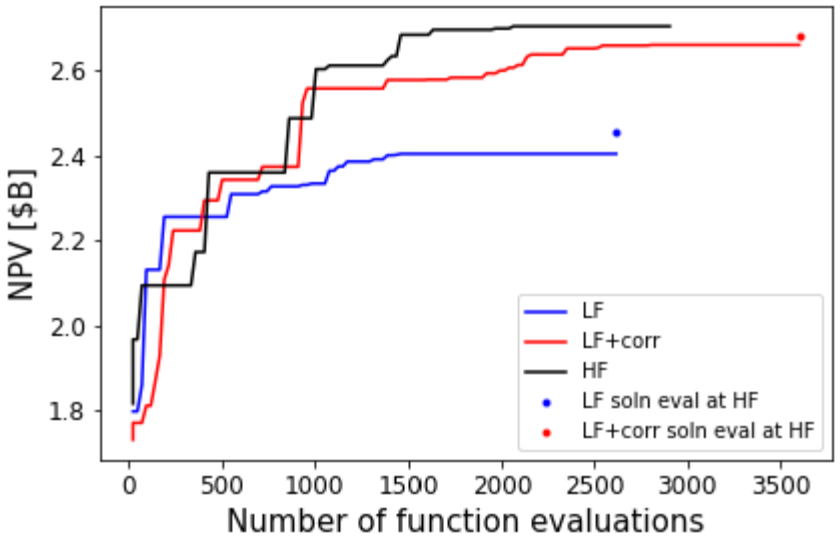}
\caption{Progress of optimizations for best-case results over all 25 runs (Case~1). Solid curves show NPV results during the course of the optimization, evaluated on HF, LF, or LF+corr model. Blue and red points denote NPV for LF and LF+corr optimal configurations evaluated on HF model.} 
\label{fig:wpo_4w_process_final}
\end{figure}

\begin{figure*}[!htb]
\begin{subfigure}{.32\textwidth}
  \centering
  \includegraphics[width=1.3\linewidth]{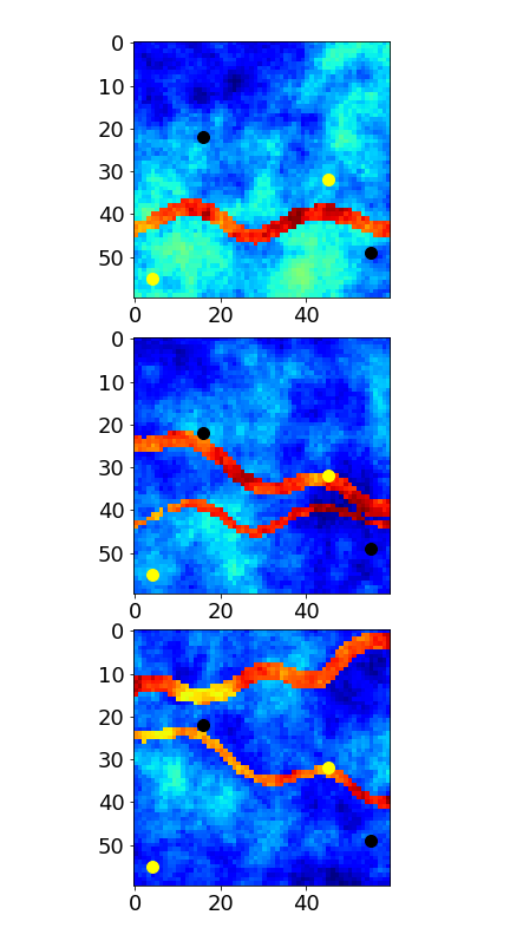}  
  \caption{LF}
  \label{fig:wpo_4w_loc_final_LF}
\end{subfigure}
\begin{subfigure}{.32\textwidth}
  \centering
  \includegraphics[width=1.3\linewidth]{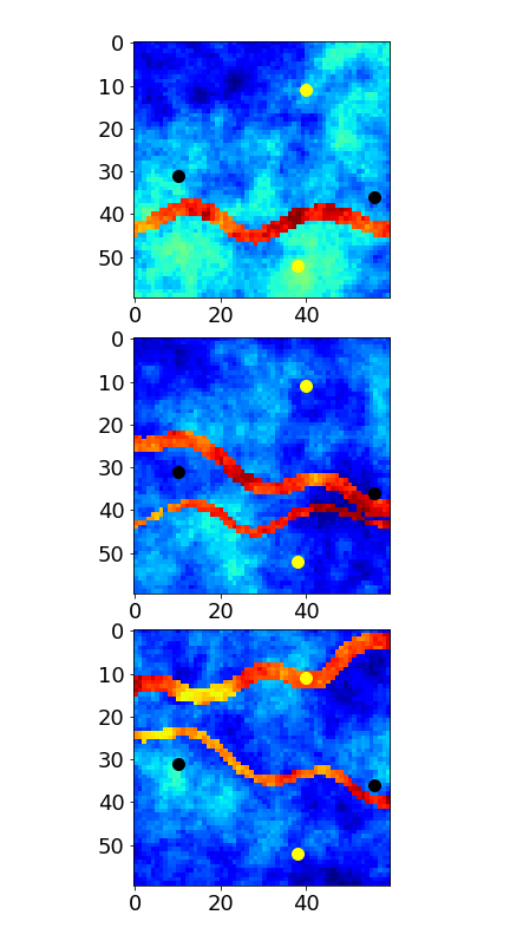}
  \caption{LF+corr}
  \label{fig:wpo_4w_loc_final_LF_CORR}
\end{subfigure}
\begin{subfigure}{.32\textwidth}
  \centering
  \includegraphics[width=1.3\linewidth]{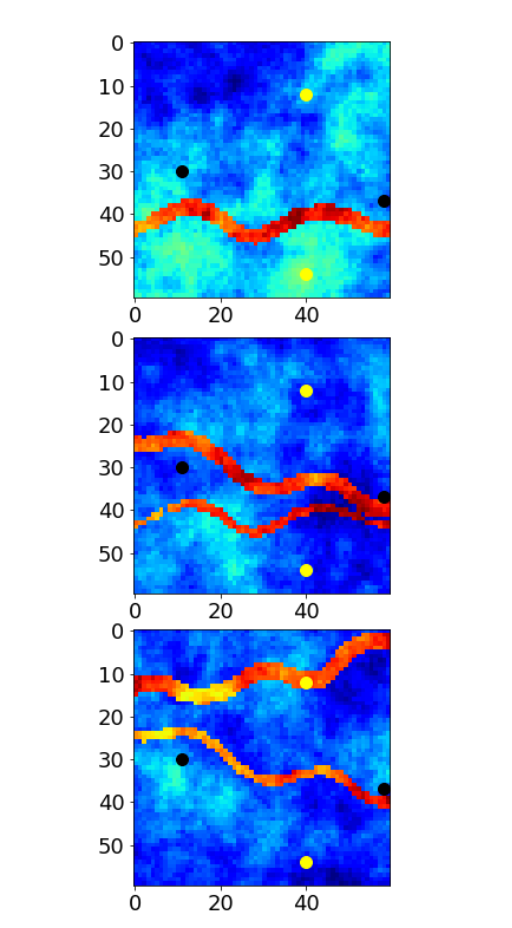}
  \caption{HF}
  \label{fig:wpo_4w_loc_final_HF}
\end{subfigure}
\caption{Optimal well configurations (production wells in yellow and injection wells in black) from the three methods for Case~1. Top row shows layer~5, middle row shows layer~10, and bottom row shows layer~15.}
\label{fig:wpo_4w_loc_final}
\end{figure*}

A perfect error model would shift all the LF results (circles) onto the 45-degree line. We see that, in most cases, the LF+corr results (stars) do fall nearer the 45-degree line than the circles. In many cases the improvement is substantial, though there are scenarios (e.g., those corresponding to the second and third-lowest HF NPVs) where the LF+corr result is less accurate than the LF result. Relative error is shown, in terms of box plots, in Fig.~\ref{fig:error_model_4w}b. Here relative error is the absolute error as a fraction of the  NPV evaluated on the HF model. In these box plots, the top and bottom of the box indicate the 75th and 25th percentile errors (these are referred to as P$_{75}$ and P$_{25}$ errors), the line inside the box indicates the median (P$_{50}$) error, and the `whiskers' outside the box show the maximum and minimum errors. The LF+corr model clearly provides lower relative error than the LF model. Specifically, the P$_{75}$ error improves from $5.5\%$ to $2.6\%$, the P$_{50}$ error from $4.7\%$ to $1.3\%$, and the P$_{25}$ error from $3.4\%$ to $0.8\%$.

During the online stage, we perform 25 optimization runs using the LF+corr model. As a reference, we also perform 25 runs with the HF model. The same initial populations are used in these runs as in the offline stage runs. In practice, because optimizations with HF models are computationally intensive, a more typical number of runs would be, e.g., three or five. We will therefore identify a subset of five HF optimization runs that might better represent the results obtained in an actual study.

Optimization results for the 25 runs using each approach are presented in Fig.~\ref{fig:wpo_4w_final}. For the LF and LF+corr runs, the optimal configurations (at the end of the run) are simulated on the HF model and NPVs are computed based on these results. Thus, direct comparisons are appropriate as all NPVs correspond to HF simulation results. For the HF optimizations, we show results from all 25 runs (open black boxes) along with a representative subset of five runs (solid black boxes). The latter correspond to the P$_{10}$, P$_{30}$, P$_{50}$, P$_{70}$, and P$_{90}$ results for the 25 HF runs. We see in Fig.~\ref{fig:wpo_4w_final} that the LF+corr treatment clearly outperforms the use of uncorrected LF models, and that it provides results that are close to those achieved using HF optimization. In particular, the NPVs for the best runs are $2.45 \times 10^9$~USD using LF models, $2.68 \times 10^9$~USD using LF+corr, and $2.70 \times 10^9$~USD using HF models. Thus, the use of the LF+corr treatment provides a 9.4\% improvement relative to LF optimization, and it is within 1\% of the result using HF optimization.

The progress of the best optimization runs using each of the three approaches, in terms of NPV versus number of flow simulations (function evaluations), is shown in Fig.~\ref{fig:wpo_4w_process_final}. We see rapid rise in NPV at early iterations, followed by slower increases as the optimizations proceed. These runs terminate based on the improvement criterion defined in Eq.~\ref{eqn:opt_termination_criteria}. The blue and red curves display NPVs evaluated using LF simulation results and the LF+corr treatment, respectively. The blue and red points at the end of the runs denote NPV evaluated on the HF model using the optimal configurations. Consistent with Fig.~\ref{fig:wpo_4w_final}, these values are $2.45 \times 10^9$~USD for LF optimization and $2.68 \times 10^9$~USD for the LF+corr treatment. Note that the difference between the point and the curve is less for LF+corr, as would be expected.

We now quantify the computational costs associated with optimizations using LF models, the LF+corr treatment, and HF models. Our assessment here follows the computational complexity approach used by Kostakis et al.~\citep{kostakis2020multifidelity}. We proceed in this manner because observed timings can vary considerably based on cluster usage, job scheduling, i/o, etc. All costs here are expressed in units of equivalent HF simulation runs. Neglecting simulator overhead, which is small compared to modeling flow dynamics in practical cases, simulation costs are approximately proportional to the number of grid blocks in the model raised to a power $\alpha \ge 1$. Thus we can write
\begin{equation}\label{eq:upsc_cost_def}
    \begin{split}
        c^l = \left(\frac{M^l}{M^h}\right)^\alpha,
    \end{split}
\end{equation}
where $M^l$ and $M^h$ are the number of cells in the LF and HF models and $c^l$ is the  computational cost of a LF run in units of HF simulation runs. Here we take $\alpha = 1$, which gives $c^l = \frac{10 \times 10 \times 10}{60 \times 60 \times 30}=\frac{1}{108}$ for our case. This is a conservative estimate for $\alpha$, and the use of $\alpha>1$ would lead to larger speedups. Upscaling costs are not included in these assessments, as these are small compared to even LF oil-water flow simulation. We also neglect the runtime cost of applying the error model, which is again very small compared to flow simulation costs. In the LF+corr costs, we do include the costs for the offline stage. These derive from the 25 optimization runs (using LF models) and the 100 HF simulation runs required to train the error model. 

The computational costs for the various approaches, along with the median and best NPV results, are presented in Table~\ref{tab:wpo_4w_final_summary}. The 25 HF optimization runs require a total of 79,080 HF flow simulations ($N_{opt}=25$, $N_p=24$, $N_{iter} \approx 132$). Optimization using uncorrected LF models provides a speedup factor of about 100 ($c_l \approx 0.01$), while that using LF+corr leads to a speedup factor of 46. The extra computation for LF+corr relative to LF derives mostly from the LF runs performed during the preliminary (offline) $N_{opt}^0=25$ optimizations. This cost could be reduced, if necessary, by performing fewer runs or by using fewer iterations per run. The NPV results in Table~\ref{tab:wpo_4w_final_summary} reiterate our observations from Fig.~\ref{fig:wpo_4w_final} -- namely, that the LF+corr treatment provides optimized NPVs that are close to those from HF optimization. Importantly, however, the computational cost associated with LF+corr is much closer to that of LF optimization than to HF optimization.

\begin{table}[htb!]
\centering
    \caption{Summary of optimization results and costs (for 25 runs) for Case~1. NPVs are in 10$^9$~USD, and optimization costs are in equivalent HF simulations}
    \begin{tabular}{|r|r|r|r|}
    \hline
    {} & {Opt Cost} & {Median NPV} & {Best NPV}  \\ \hline
    HF   & 79,080  & 2.49 & 2.70 \\
    LF & 772  & 2.35 & 2.45 \\
    LF+corr & 1722  & 2.44 & 2.68 \\
    \hline
    \end{tabular}%
    \label{tab:wpo_4w_final_summary}%
\end{table}%

\begin{figure}[!htb]
\centering
\includegraphics[width = 0.48\textwidth]{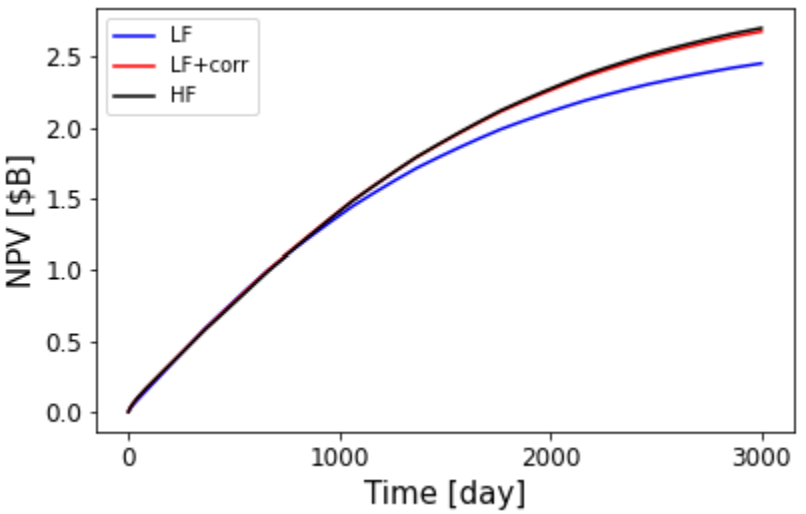}
\caption{Evolution of NPV with time for the best run for each of the three optimization approaches (Case~1).}\label{fig:wpo_4w_npv_time}
\end{figure}

\begin{figure}[!htb]
\centering
\includegraphics[width = 0.48\textwidth]{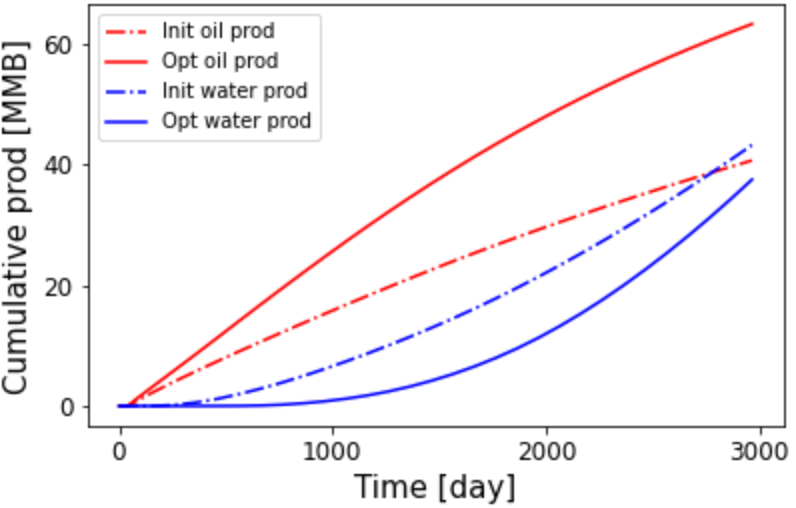}
\caption{Cumulative field-wide oil and water production for initial and optimal solutions using LF+corr procedure (Case~1). Initial results are for the best configuration in the first optimization iteration.} 
\label{fig:wpo_4w_cum_rat}
\end{figure}

We now present the optimal well configurations and corresponding simulation results for the three optimization treatments. The well locations for the best run with each approach are shown in Fig.~\ref{fig:wpo_4w_loc_final}. Here the yellow circles indicate producers and the black circles injectors. The wells are vertical and penetrate the full model, so their areal ($x$-$y$) locations are the same in all layers. High-permeability channel features appear in different locations in the three layers shown in Fig.~\ref{fig:wpo_4w_loc_final}, which renders the flow problem truly 3D and thus more challenging. The well locations are very similar (but not identical) for the LF+corr and HF optimizations. Because WPO problems tend to have many local optima, it is common to observe different configurations that correspond to similar objective function values. In this case, however, we achieve very similar configurations (and NPVs) with the two approaches. The well locations from LF optimization are quite distinct from those for the other two methods.

\begin{figure*}
\begin{subfigure}{.32\textwidth}
  \centering
  \includegraphics[width=1.2\linewidth]{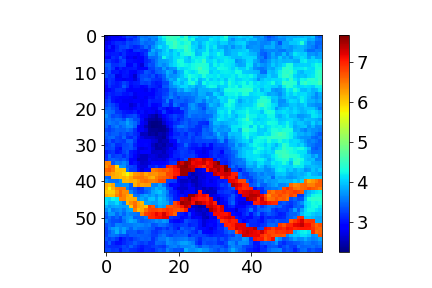}  
  \caption{Layer~5}
  \label{fig:res_model_8w_layer5}
\end{subfigure}
\begin{subfigure}{.32\textwidth}
  \centering
  \includegraphics[width=1.2\linewidth]{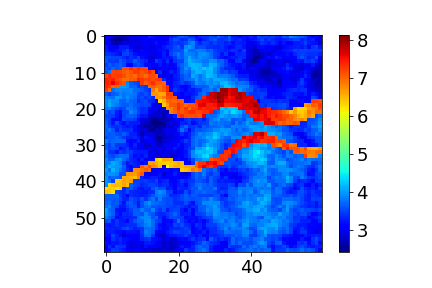}
  \caption{Layer~15}
  \label{fig:res_model_8w_layer15}
\end{subfigure}
\begin{subfigure}{.32\textwidth}
  \centering
  \includegraphics[width=1.2\linewidth]{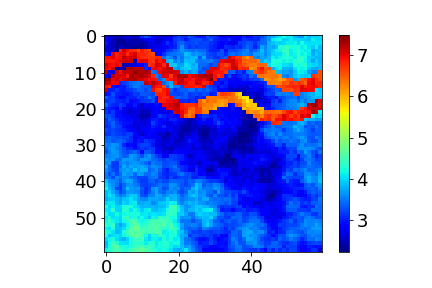}
  \caption{Layer~20}
  \label{fig:res_model_8w_layer20}
\end{subfigure}
\caption{Permeability field ($\log_e k_x$, with $k_x$ in millidarcy) for three layers of the HF geomodel (Case~2).}
\label{fig:res_model_8w}
\end{figure*}

\begin{figure*}[!htb]
\begin{subfigure}{.5\textwidth}
  \centering
  \includegraphics[width=0.95\linewidth]{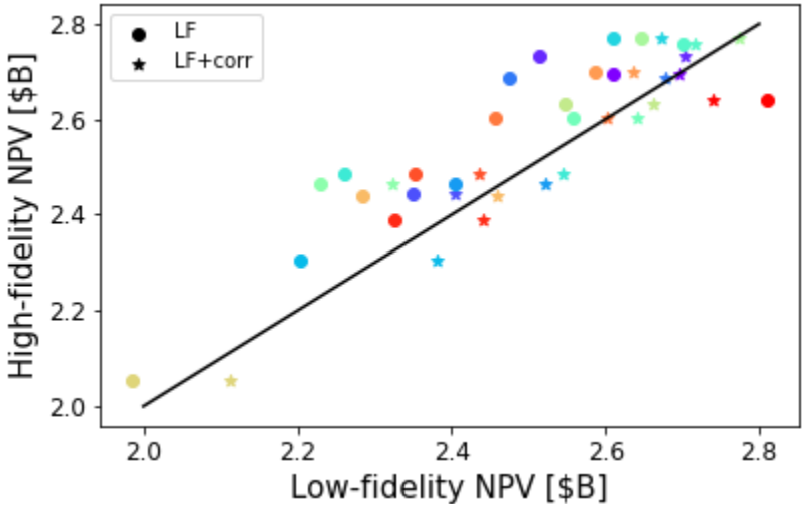}
  \caption{Cross-plots of NPV evaluated on HF, LF, and LF+corr models}
  \label{fig:error_model_8w_dot}
\end{subfigure}
\begin{subfigure}{.5\textwidth}
  \centering
  \includegraphics[width=0.95\linewidth]{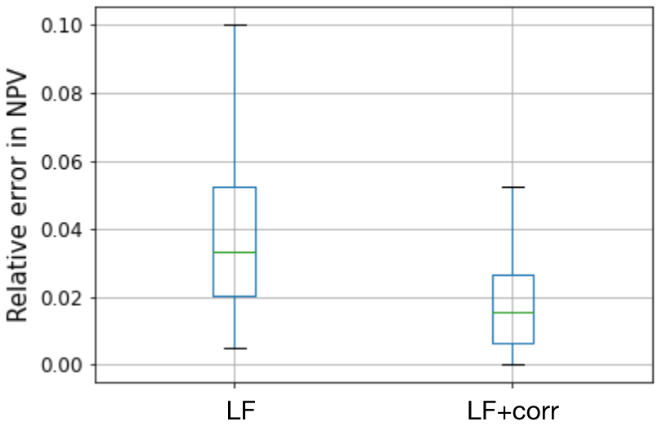}
  \caption{Box plots of relative error}
  \label{fig:error_model_8w_box}
\end{subfigure}
\caption{Test sample results for Case~2.}
\label{fig:error_model_8w}
\end{figure*}

The evolution of NPV with time for the three optimization procedures (for the best run with each method, simulated at HF) is presented in Fig.~\ref{fig:wpo_4w_npv_time}. The close correspondence between the HF and LF+corr optimization results is again evident. Results for field-wide cumulative oil and water production as a function of time are displayed in Fig.~\ref{fig:wpo_4w_cum_rat}. The optimized results are for the best LF+corr run, and the initial production curves correspond to the best solution at the first iteration of this run. Thus we see that the optimized well locations lead to much more oil production and less water production than in the initial solution.

\begin{figure}[!htb]
\centering
\includegraphics[width = 0.48\textwidth]{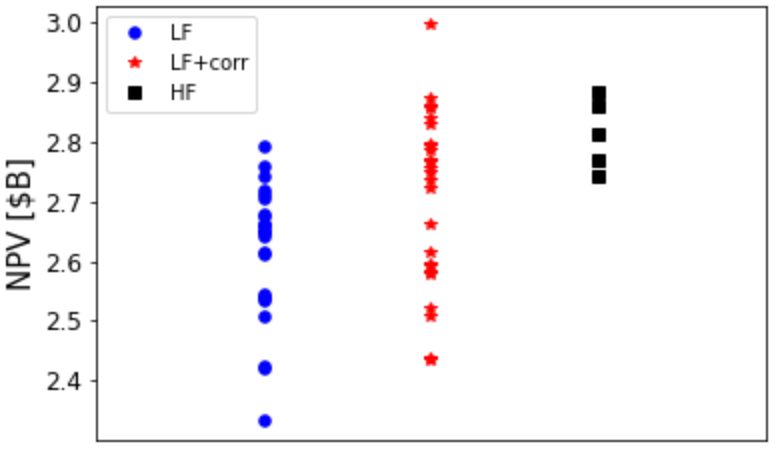}
\caption{Optimization results for Case~2. All NPVs are evaluated on the HF model.} \label{fig:wpo_8w_final}
\end{figure}

\begin{table}[htb!]
\centering
    \caption{Summary of optimization results and costs for Case~2. NPVs are in 10$^9$~USD, and optimization costs are in equivalent HF simulations. HF costs are for five optimization runs, while LF and LF+corr costs are for 25 runs}
    \begin{tabular}{|r|r|r|r|}
    \hline
    {} & {Opt Cost} & {Median NPV} & {Best NPV}  \\ \hline
    HF (5 runs)   & 11,568 runs  & 2.81 & 2.89 \\
    LF & 624 runs  & 2.65 & 2.79 \\
    LF+corr & 1402 runs  & 2.76 & 3.00 \\
    \hline
    \end{tabular}%
    \label{tab:wpo_8w_final_summary}%
\end{table}%

\begin{figure*}[!htb]
\begin{subfigure}{.32\textwidth}
  \centering
  \includegraphics[width=1.3\linewidth]{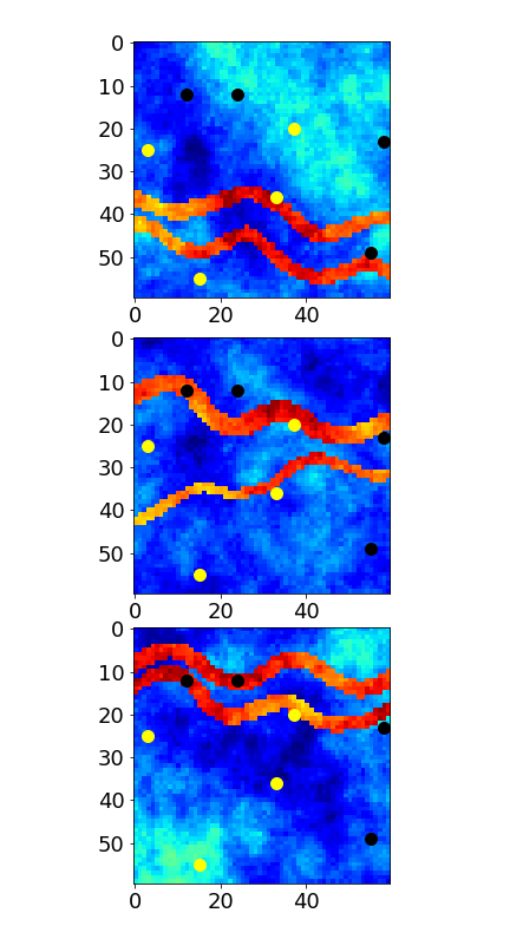}  
  \caption{LF}
  \label{fig:wpo_8w_loc_final_LF}
\end{subfigure}
\begin{subfigure}{.32\textwidth}
  \centering
  \includegraphics[width=1.3\linewidth]{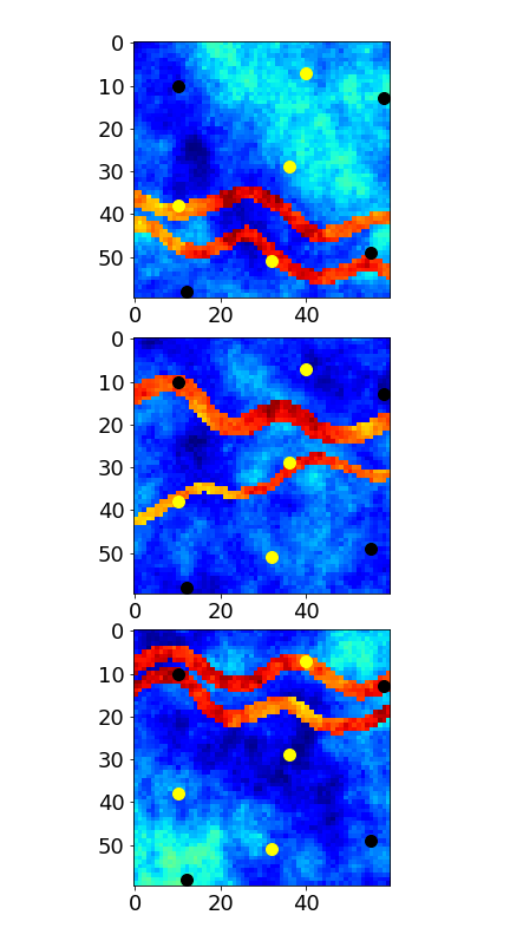}
  \caption{LF+corr}
  \label{fig:wpo_8w_loc_final_LF_CORR}
\end{subfigure}
\begin{subfigure}{.32\textwidth}
  \centering
  \includegraphics[width=1.3\linewidth]{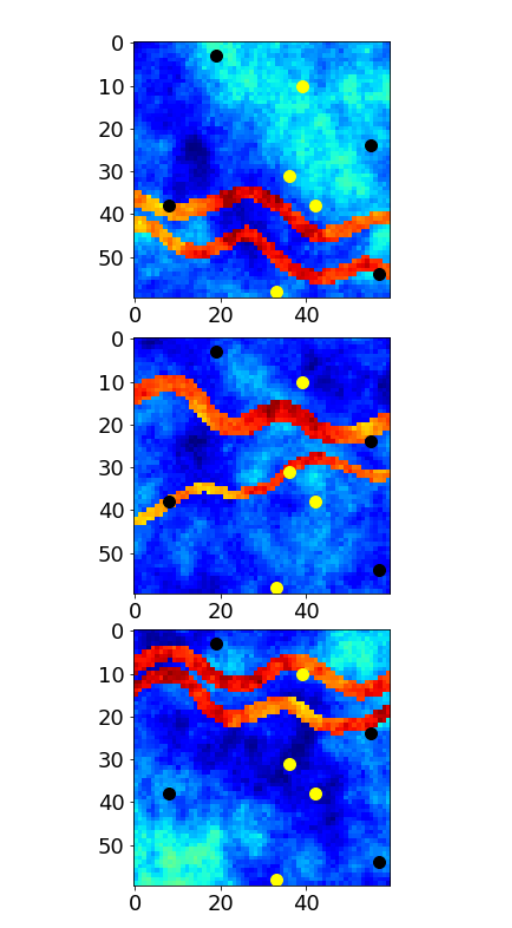}
  \caption{HF}
  \label{fig:wpo_8w_loc_final_HF}
\end{subfigure}
\caption{Optimal well configurations (production wells in yellow and injection wells in black) from the three methods for Case~2. Top row shows layer~5, middle row shows layer~15, and bottom row shows layer~20.}
\label{fig:wpo_8w_loc_final}
\end{figure*}

\subsection{Case~2: four injectors and four producers}
\label{sec:case2}

The setup for Case~2 differs somewhat from that for Case~1. We now use a different geomodel (a new realization from the same geological scenario as in Case~1), and we consider $N_w=8$~wells (four injectors and four producers) instead of $N_w=4$ as in Case~1. In addition, the simulation time frame is now 2000~days, as opposed to 3000~days for Case~1 (because there are now more wells, the reservoir can be swept in less time). Three layers of the (HF) bimodal channelized system considered in this case are shown in Fig.~\ref{fig:res_model_8w}. The channel locations clearly differ from those in Case~1 (Fig.~\ref{fig:res_model_4w}).

In the offline stage, we perform 25 LF optimization runs and then select 150 different well configurations from the full set of results (using the clustering procedure) to train the error model. We divide these 150 samples into 130 for training and 20 for testing. Test results for the error model for Case~2 are presented in Fig.~\ref{fig:error_model_8w}. Consistent with the Case~1 results in Fig.~\ref{fig:error_model_4w}, we again see clear improvement on a case-by-case basis (compare circles and stars in Fig.~\ref{fig:error_model_8w}a) as well as on a quantile basis (box plots in Fig.~\ref{fig:error_model_8w}b). In this case the P$_{50}$ error improves from 3.3\% (for LF models) to 1.6\% (for LF+corr).

We now compare optimization performance for the three methods for Case~2. Optimized NPV results are presented in Fig.~\ref{fig:wpo_8w_final}. Note that in this case we perform only five HF optimization runs (this is consistent with practical WPO using HF models). Thus there are only five black squares in Fig.~\ref{fig:wpo_8w_final}. As in Case~1, we again perform 25 LF and LF+corr optimization runs. For these runs, the optimal configurations are again simulated on the HF model, so direct comparison of NPVs in Fig.~\ref{fig:wpo_8w_final} is appropriate. It is immediately apparent that the best LF+corr NPV exceeds that of the best HF optimization result in this case. 

Summary results for median and best-case NPV and cost are presented in Table~\ref{tab:wpo_8w_final_summary}. The HF costs are for five optimization runs (they are more than a factor of five less than the HF cost in Table~\ref{tab:wpo_4w_final_summary} because fewer optimization iterations are required in Case~2), while the LF and LF+corr costs are for 25 runs. We see that, even with five times as many runs, LF+corr still provides a speedup of over a factor of eight relative to HF. In addition, because many runs can be performed inexpensively with this approach, we achieve a best-case NPV that exceeds that of HF optimization by 3.7\%. This illustrates a major advantage of the LF+corr treatment, namely, its ability to perform a large number of optimization runs, which enables a much broader overall search, while still achieving significant computational savings relative to HF optimization.

The well locations corresponding to the optimal solutions from the three approaches are shown in Fig.~\ref{fig:wpo_8w_loc_final}. In contrast to the Case~1 configurations (Fig.~\ref{fig:wpo_4w_loc_final}), we now observe clear differences between the HF and LF+corr well locations. These two solutions do share some commonalities, however, as they both contain injection wells near the model boundaries, with production wells toward the interior of the model. This is an intuitive solution, though the detailed locations are of course strongly impacted by the channels and permeability variations. The configuration from the LF optimization is somewhat different, as two of the injectors are near the middle of the model. This configuration provides a lower NPV than those in Fig.~\ref{fig:wpo_8w_loc_final}b and c, though it does outperform many of the HF and LF+corr solutions, as is evident from the results in Fig.~\ref{fig:wpo_8w_final}. 

\begin{figure}[!htpb]
\centering
\includegraphics[width = 0.48\textwidth]{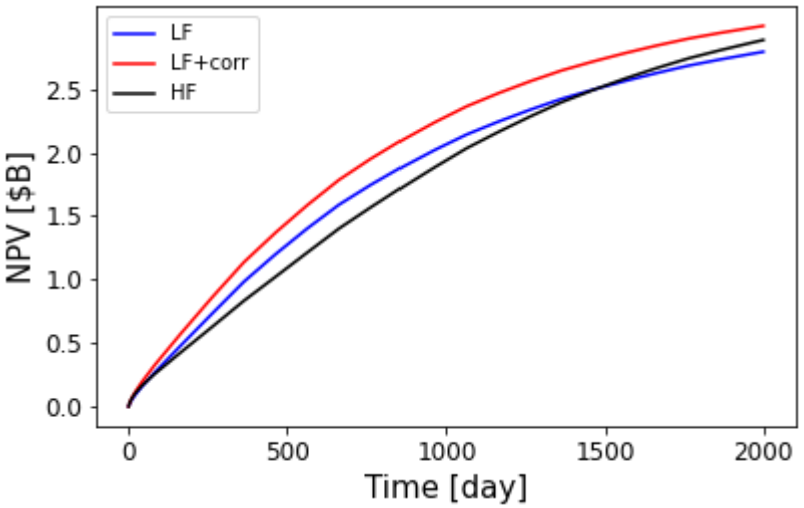}
\caption{Evolution of NPV with time for the best run for each of the three optimization approaches (Case~2).}
\label{fig:wpo_8w_npv_time}
\end{figure}

In Fig.~\ref{fig:wpo_8w_npv_time}, we present the evolution of NPV with time for the optimal configurations from the three procedures. All results are for the best case, simulated on the HF model. We see that the LF+corr approach provides the highest NPV at all times. It is also evident that NPV plateaus earlier in this case than in Case~1 (Fig.~\ref{fig:wpo_4w_npv_time}). This results from the use of eight wells in this case. 

\section{Concluding remarks}
\label{sec:summary}

In this paper, we introduced an optimization strategy that entails the use of low-fidelity (highly upscaled) simulation models in combination with tree-based machine-learning error correction. The procedure was applied for well placement optimization in heterogeneous 3D geomodels. With this treatment, LF models for each candidate well configuration are constructed using a global transmissibility upscaling procedure. During the offline or preprocessing stage, we use a differential evolution (DE) optimizer, with all simulations performed on LF models, to establish a set of 100--150 well configurations that is appropriate to use for training. Both random forest and light gradient boosting machine models are then trained to estimate the error in objective function value, here taken to be NPV, that results from performing simulation with LF (rather than HF) models. In the online or runtime stage, DE optimization is again performed, with NPV for each candidate well configuration estimated from LF simulation results and the error models.

The optimization framework was then used for two example cases involving the placement of different numbers of wells. Three different procedures were applied to enable the assessment of our new treatments -- optimization using LF models, optimization using HF models, and optimization using LF models with the error correction (this approach is referred to as LF+corr). In the first case, involving two injectors and two producers, we performed 25 optimization runs with each method. The LF+corr procedure provided an overall speedup factor of 46 relative to HF optimization, and it gave a best-case NPV that was within 1\% of the best HF optimization result. The LF+corr best-case NPV exceeded that achieved using uncorrected LF models by 9.4\%. In the second case, involving eight wells, we conducted 25 LF and LF+corr optimization runs, but only five HF runs (this is a more realistic number of HF runs given the expense of these computations). In this case the LF+corr procedure gave an overall speedup factor of 8.3 relative to HF optimization, and it resulted in a best-case NPV that exceeded the best-case HF NPV by 3.8\%. This result illustrates a key advantage of our procedure; i.e., because it is much faster than optimization using HF models, a large number of runs can be conducted, which is beneficial with stochastic search algorithms.

There are a number of directions that could be targeted in future research in this area. The overall framework should be extended to enable treatment of a wider range of problems. These could include three-phase flow models, cases with horizontal or deviated wells, systems with more constraints, and studies involving geological uncertainty. In the latter case multiple geological realizations would need to be considered, and different error modeling approaches should be evaluated. These could entail treatments where the goal is to correct results on a realization-by-realization basis, and approaches where only the expected NPV (averaged over all realizations) is corrected. It may also be useful to further tune some of the detailed treatments within the framework. Specifically, the development of guidelines for determining the optimal number of offline optimization runs and training samples, and further exploration of the impact of the hyperparameters associated with the error models, might be pursued. Finally, additional ML methods, including other tree-based approaches (e.g., Extra Trees and Xgboostm) and shallow artificial neural networks, should be considered for modeling the error resulting from LF simulation.

\begin{acknowledgements}
We are grateful to the Stanford Center for Computational Earth \& Environmental Sciences for providing computational resources. We thank Dylan Crain for providing the 3D geomodels and upscaling code used in this study. We are also grateful to Yimin Liu for providing the ResSimAD simulator and to Oleg Volkov for his assistance with the Stanford Unified Optimization Framework. 
\end{acknowledgements}

\small {\noindent {\bf Funding information} The authors received financial support from the industrial affiliates of the Stanford Smart Fields Consortium.}

\small {\noindent {\bf Competing interests} We declare that this research was conducted in the absence of any commercial or financial relationships that could be construed as a potential conflict of interest.}

\small {\noindent {\bf Data availability} Inquiries regarding the data should be directed to Haoyu Tang, hytang@stanford.edu.}

\bibliographystyle{elsarticle-num-names}  
\bibliography{ref}

\end{document}